# ELECTRONIC PROPERTIES OF EUMELANIN MONOMERS AND DIMMERS

*M. M. Falla Solórzano, L. E. Bolívar Marínez, S. T. Pérez Merchancano*
*Departamento de Física, Universidad del Cauca calle 5 # 4-70*
*Popayán, Cauca Colombia.*

The melanin is a group of biological pigments commonly found in living beings, it can be classified in three groups: eumelanin human beings, pheomelanins in animals, and allomelanins in the vegetal kingdom. There is a special interest in the eumelanin because this biopolymer exhibits the typical properties of a semiconductor. Eumelanin is also responsible of the main cellular photoprotection mechanisms in the human beings. The exact structural pattern of eumelanin is not completely known yet, the planar molecules 5, 6-indolquinine and its reduced forms semiquinone and hydroquinone outline the greater part of the biological pigment. In this work the structural, electronic and optical properties of monomers in various charged states (0, -1) and dimmers of eumelanin in vacuum, are found for the neutral state of charge, by means of the Semiempirical methods MNDO, AM1, PM3 and ZINDO/S-CI.

## 1. INTRODUCTION

In the past three decades, the theoretical and experimental studies of the optical properties of the eumelanin have shown that it can behave like an organic semiconductor [1]; it presents hollows or electrons conduction and a gap that oscillates between 0, 23 and 2.4eV [2]. There are points in favor of the organic semiconductors, such as they are easily manufactured, have very good mechanical properties and above all low production costs [3].

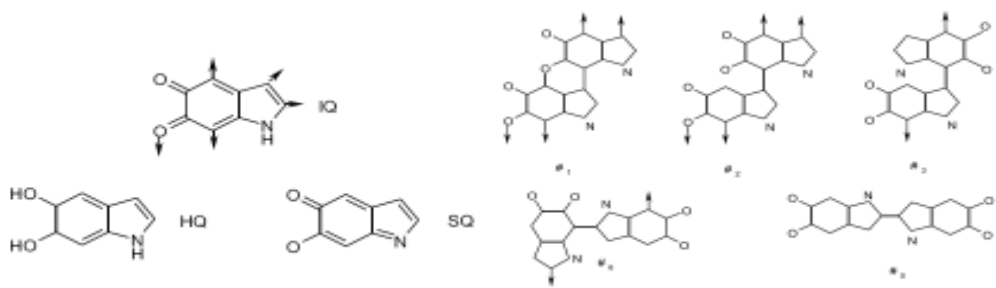

**Fig.1** Schematic representation of eumelanin monomers and dimmers. The arrows indicate the polymerization directions.

The objective of this work is the structural and optical characterization of monomers and dimers of the eumelanin (fig.1), to determine the structures that approach the reality, the polymerization form that can be presented with greater probability and if the monomers can behave like semiconductor.

## 2. METHODOLOGY

The structural characterization of monomers was carried out in vacuum and aqueous environments, the dimers (HQ-HQ, IQ-IQ, SQ-SQ) in all the possible linking structures were characterized in vacuum environment only, by means of the semiempirical methods MNDO (Modified Neglect of Differential Overlap), AM1 (Austin method 1), and PM3 (Parametric method 3) of the NDO family. Given the poor quality in the treatment of the excited states and electron correlation which is inherent to the NDO family methods, the simulation of the electronic or ultraviolet spectra will be obtained with the Zerner's spectroscopic version of the Intermediate Neglect of Differential Overlap in configuration interactions mode: ZINDO/S CI, with a 3 occupied and 3 unoccupied orbital criterion. The optical characterization of monomers was carried out in vacuum and aqueous environments obtaining besides the theoretical electronic absorption spectrum UV, many important characteristics such as the point were the absorption begins, the maximum absorption peaks, and the most relevant electronic transitions.

## 3. RESULTS

Heat of formation of monomers in neutral state and of charge -1 in vacuum and aqueous environment are shown in table 1, for the vacuum environment the best values are provided by the MNDO method in aqueous environment the best values are given by the AM1 method.

**Table 1.** Heat of formation of the monomers in neutral state and charge -1 in vacuum and aqueous environment.

| Semiempirical Method | Charge state | HQ | | IQ | | SQ | |
|---|---|---|---|---|---|---|---|
| | | Vacuum | Water | Vacuum | Water | Vacuum | Water |
| MNDO | 0 | -49,5221 | -1541,7124 | -8,2171 | -1438,0779 | -5,2282 | -1497,1006 |
| | -1 | -64.4115 | -1578,5655 | -53.8723 | -1505,7654 | -59.6971 | -1576,3450 |
| AM1 | 0 | -29,3083 | -1662,2673 | 12,0015 | -1603,3991 | 23,4897 | -1591,5915 |
| | -1 | -41.0985 | -1684,8606 | -36.4485 | -1687,6908 | -34.7580 | -1694,6666 |
| PM3 | 0 | -44,2146 | -1449,0429 | -9,1559 | -1349,6219 | 4,3981 | -1389,8969 |
| | -1 | -55.6256 | -1491,7891 | -57.7734 | -1435,1008 | -53.6433 | -1497,7307 |

Heat of formation values show that the most favorable state of charge for the formation of the molecular structures here studied is the -1, which implies that these molecules behave as electron acceptors. The structures obtained from the geometrical optimizations with the methods AM1 and MNDO were then studied with the ZINDO/S CI, to therefore obtain the values of wavelength and the oscillator strength that conform the UV spectrum, this calculation were performed also in vacuum and aqueous environments (figure 2). In the fig 2a we can observe that the HQ monomers begin to absorb in the 152 nm region and that the maximum absorption is located around the 221 nm, for the other structures the results are similar, the absorption region for the neutral state of charge is therefore located around the 5.59 eV implying that this structure tend to behave like an isolator. For the monomers in the -1 charge state (fig 2b), the optical gap is found to diminish to the 1.9 eV value showing the tendency of this structures to behave like a semiconductor when they are ionized. The most prominent optical transitions appear to be of the H→L+1, H -1→L+2 of the sigma – pi type implying that this molecules are not planar given that hydrogen and oxygen atoms are displaced from the main structural plane.

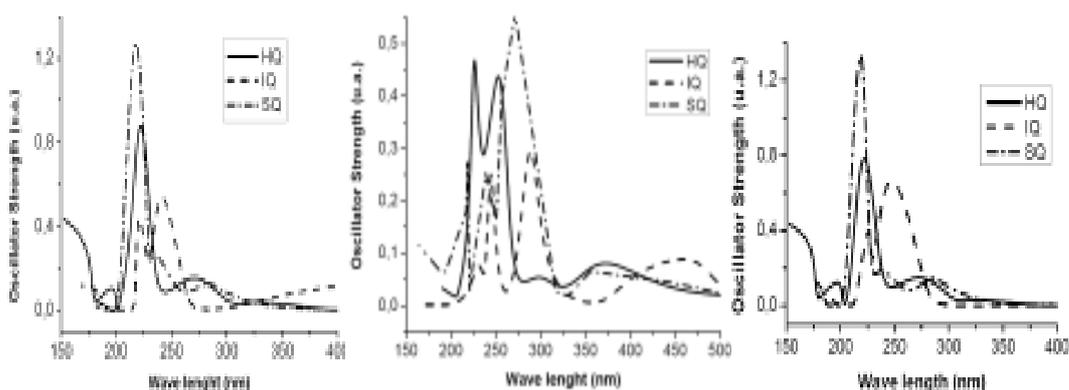

**FIG. 2 (a)** Absorption spectrum UV of the monomers in vacuum in state of charge 0, **(b)** Absorption spectrum UV of the monomers in vacuum in state of charge -1, **(c)** Absorption spectrum UV of the monomers in water in state of charge 0.

In the aqueous environment (fig 2b) the simulations of the electronic absorption spectra show that the HQ molecule begins to absorb in the range of the 237 nm, value that corresponds to nearly 5.23 eV. This results are very similar to those found previously for the monomers in the vacuum and neutral charge environment, given this we can affirm that in general this molecules tend to behave as isolators with no dependence at all of the environment in witch are studied (aqueous or vacuum).

The dimmers were constructed joining monomers of the same kind (HQ-HQ, IQ-IQ, SQ-SQ) taking into account the linking mechanism shown in fig 1. The results of the geometric optimizations (table 2) show that the best values for the HQ-HQ, SQ-SQ structures are obtained with the MNDO and for the IQ-IQ ones with the PM3 method.

**Table 2.** Heat of formation of dimmers in neutral state in vacuum

| Polymerization states | HQ | | | IQ | | | SQ | | |
|---|---|---|---|---|---|---|---|---|---|
| | MNDO | AM1 | PM3 | MNDO | AM1 | PM3 | MNDO | AM1 | PM3 |
| 1 | -77,238 | -24,7314 | -63,977 | | | | 9,301 | 78,706 | 30,753 |
| 2 | -99,375 | -58,2197 | -87,025 | -14,5296 | 27,295 | -15,090 | -7,652 | 50,469 | 14,950 |
| 3 | -99,375 | -58,2197 | -87,025 | -14,5296 | 27,295 | -15,090 | -7,652 | 50,469 | 14,950 |
| 4 | -101,448 | -61,7539 | -90,576 | -15,7330 | 27,024 | -16,952 | -8,729 | 50,903 | 5,5540 |
| 5 | -103,158 | -60,1740 | -92,823 | -16,2538 | 29,187 | -18,096 | -14,002 | 46,411 | 5,6604 |

The polymerization state # 1 is not possible for the IQ-IQ structure given that the valence limit is exceeded. The polymerization form # 5 for all the dimmers can appear in spontaneous form in the nature given that this is structure with the lowest energy of conformation, yet this polymerization mechanism is characterized with the absence of a defined growth direction; this means that in the biopolymer chains of melanin the structures created with this mechanism can appear in the terminal point which effectible closes the chain preventing a further polymeric develop.

## 4. CONCLUSIONS

The geometric optimization of the monomer in the different states of charge and of the dimmers in the vacuum environment provided the best values for the heat of formation using the MNDO method excluding the IQ-IQ dimmer whose best values were given by the PM3 method. For the structures in aqueous environment the best values were given by the AM1 method. This results show that the monomers in charged state can exist in spontaneous form in the nature and that they are charge acceptors performing cellular functions as free radical catchers in diseases like the Parkinson's. The dimmers formed from the joining of two identical monomers favor the # 5 mechanism given that this is the lowest energy values conformation; nevertheless this form is not apt for the formation of long chains due to its presence in the terminal parts of the chain. Because of the previous condition the polymerization process favors the #4 linking mechanism.

The theoretical UV spectra of the monomers in neutral state and aqueous and vacuum environments shown that this structures exhibit an isolator behavior independently of the medium because of the high optical gap values. For the monomers in the -1 state of charge the values of the optical gap change and the structures can behave like a semiconductor.